\documentclass[3p,twocolumn]{elsarticle}

\usepackage{amssymb}
\usepackage{amsmath}
\usepackage{lineno}
\journal{Chaos Solitons and Fractals}

\begin{document}


\begin{frontmatter}

\title{Competition of alliances in a cyclically dominant eight-species population}

\author[label1]{Junpyo Park}
\cortext[mail1]{Email addresses: junpyopark@khu.ac.kr; xiaojiechen@uestc.edu.cn; szolnoki.attila@ek-cer.hu}
\author[label2]{Xiaojie Chen}
\author[label3]{and Attila Szolnoki\corref{ca}}
\cortext[ca]{Corresponding author}
	
\address[label1]{Department of Applied Mathematics, Kyung Hee University, Yongin 17104, Republic of Korea}
\address[label2]{School of Mathematical Sciences, University of Electronic Science and Technology of China, Chengdu 611731, China}
\address[label3]{Institute of Technical Physics and Materials Science, Centre for Energy Research, P.O. Box 49, H-1525 Budapest, Hungary}

\begin{abstract}
In a diverse population, where many species are present, competitors can fight for surviving at individual and collective levels. In particular, species, which would beat each other individually, may form a specific alliance that ensures them stable coexistence against the invasion of an external species. Our principal goal is to identify those general features of a formation which determine its vitality. Therefore, we here study a traditional Lotka-Volterra model of eight-species where two four-species cycles can fight for space. Beside these formations, there are other solutions which may emerge when invasion rates are varied. The complete range of parameters is explored and we find that in most of the cases those alliances prevail which are formed by equally strong members. Interestingly, there are regions where the symmetry is broken and the system is dominated by a solution formed by seven species. Our work also highlights that serious finite-size effects may emerge which prevent observing the valid solution in a small system.
\end{abstract}


\end{frontmatter}

\section{Introduction}
\label{intro}

If a species is stronger and beats the other one then how can the ``weaker'' species survive? This is a fundamental problem of ecology to explain the amazing biodiversity of life \cite{sigmund_93,bascompte_98}. A possible explanation could be the so-called intransitive or cyclically dominated relation among competing species where a third (or more) species beat the predator species, hence establishing a delicate balance among all competitors. In the simplest case, these relations remind the well-known rock-scissors-paper game and this type of interaction can really be observed among animals, plants, or even among bacterias \cite{sinervo_n96,burrows_mep98,kerr_n02,cameron_jecol09,ruifrok_tpb15,garde_rsob20,liao_ncom20}. Evidently, a three-member loop can be easily enlarged for more participants, as it was made in the extended Lotka-Volterra model \cite{frachebourg_pre96,esmaeili_pre18,baker_jtb20,bayliss_pd20}. Motivated by the complexity of such interactions, researchers have studied related models actively over the last decade and several interesting observations have been made \cite{durney_pre11,cazaubiel_jtb17,choi_epjb17,roman_jtb16,szabo_pre07}. 

It would be almost impossible to sum all important observations, such as reported in Refs.~\cite{park_c22,yoshida_srep22,tainaka_ei21,serrao_epjb21,park_c18c,nagatani_srep18, mobilia_g16}. Instead, the reader is referred to specific review papers about the milestones along this research path \cite{szolnoki_jrsif14,szabo_pr07,dobramysl_jpa18,szolnoki_epl20}. Nevertheless, it is also worth noting here that the mentioned cyclic or intransitive relation among participants should not necessarily be defined by a Lotka-Volterra-type microscopic rule, but could be the result of collective behavior among competing strategies in evolutionary game models \cite{hauert_s02,szolnoki_csf20b,palombi_epjb20,szolnoki_pre17,canova_jsp18,szolnoki_prx17,tao_yw_epl21,szolnoki_pre10b,liu_lj_rsif22}

The most exciting question is whether we can predict the direction of evolution based on the food-web that defines the microscopic dynamics? Can we identify general principles which may inform us about the vitality of an alliance? An early observation was, for instance, that when two three-member cycles fight then the one, where the inner invasion is faster, is more viable \cite{perc_pre07b}. The picture, however, is more subtle, because a faster general rotation does not necessarily provide an advantage if the members of the trio are unequal. In the latter case, the triplet becomes vulnerable no matter the average of inner invasion rates is higher than in the rival triplet which is formed by less aggressive, but equally strong partners \cite{blahota_epl20}. Notably, heterogeneous invasion rates may emerge temporarily or locally  \cite{szolnoki_srep16b,szolnoki_pre16}. Or, the extension of the original food-web by adding a reverse link can also change a stable solution \cite{bazeia_csf20}. Furthermore, the number of members could also be a decisive factor because a smaller loop is generally stronger than a large one \cite{de-oliveira_csf22}.

In this work we introduce a minimal eight-species model where two four-member loops fight for space. The interaction between these quartets practically establishes a traditional Lotka-Volterra circle of eight species, which can be described by an invasion rate. Not to break the balance between the quartets, we assume that the inner invasion is uniform and equally strong in both formations. The only difference between them is we introduce two short-cuts in one of the loops which generates extra interactions within the related alliance. Our major question is how the stability of alliances change due to this extension and can we identify new solutions which would be hidden otherwise? It could also be interesting whether the previously established principles, observed for simpler systems formed by trials and duets, remain valid when we apply them to alliances formed by larger groups.

\section{Alliances formed at different levels}
\label{def}

To make our observations comparable with previous works \cite{intoy_pre15,lutz_jtb13,mir_pre22,szabo_pre08,brown_pre19,park_c19b}, we assume that species are distributed on an $L \times L$ square lattice where periodic boundary conditions are applied. Each node is occupied by one of the species which are denoted by $X_i$ where indexes are from $i=0$ to 7. During an elementary step, we select a neighboring pair nodes. If they are occupied by the same species or represent species which are neutral then nothing happens. Otherwise, an invasion happens with a specific probability and predator species invades prey species. The microscopic rules are defined in the following way:
\begin{eqnarray}
	&X_iX_{i+1} \xrightarrow{\gamma} X_iX_i\\
	&X_iX_{i+2} \xrightarrow{\alpha} X_iX_i\\
	&X_2X_6 \xrightarrow{\beta} X_2X_2, X_4X_0 \xrightarrow{\beta} X_4X_4,
\end{eqnarray}
where $i+1$ and $i+2$ are calculated in cyclic manner. The parameters $\alpha, \beta$ and $\gamma$ define the probabilities of successful invasions between the involved predator-prey neighbors.
\begin{figure}[h!]
	\centering
	\includegraphics[width=7.5cm]{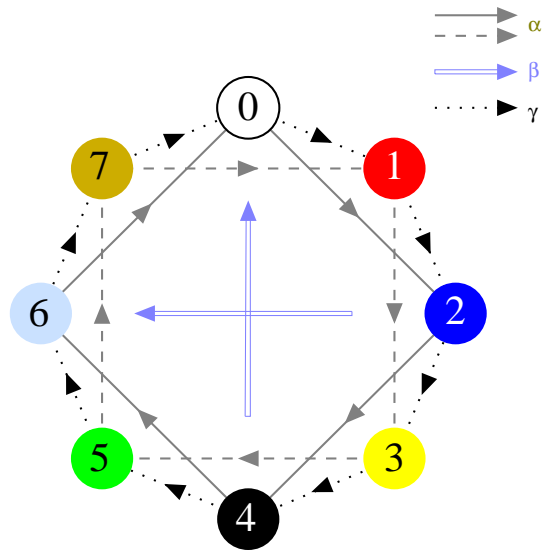}
	\caption{Food-web of competing species. In the basic model eight species are invade cyclically each other with probability $\gamma$, similar to the extended Lotka-Volterra model. Additionally, we introduce a cyclic inner invasion among odd and separately among even species with probability $\alpha$. To break the symmetry among the quartets, we also introduce an inner invasion between species ``2'' and ``6'' and between species ``4'' and ``0'' with probability $\beta$. Importantly, we use the color-code of species in later figures where spatial distributions are presented.}\label{foodweb}
\end{figure}

For a deeper insight about the model, in Fig.~\ref{foodweb} we show the food-web of competing species. Our first note is about the biggest loop where all species are involved, according to the extended Lotka-Volterra type system. Here every species is simultaneously a predator and a prey of another one, hence establishing a possible solution. The invasion rate is $\gamma$ for all interactions here. Another solution is formed by ``1'', ``3'', ``5'', and ``7'' species who invade each other cyclically with probability $\alpha$. For later reference, we denote this alliance as $A^4_{1,3,5,7}$. A similar $A^4_{0,2,4,6}$ quartet is formed among species   ``0'', ``2'', ``4'' and ``6''.  Additionally, we introduce an extra chance of invasions among group members here. In particular, species ``2'' invades species ``6'' and species ``4'' invades species ``0'' with probability $\beta$. In this way, there are no neutral pairs in the mentioned quartet anymore. As a consequence, the average inner invasion is augmented among the four species, which could be a support to their viability. Interestingly, however, there are many other possible solutions in this system. For example, the new invasions among even-numbered species offer the chance for trials to emerge: $A^3_{0,2,4}$ or $A^3_{0,2,6}$ can work as a rock-scissors-paper-type solution. Beside the mentioned formations, there are $A^5$ (five-), $A^6$ (six-), or $A^7$ seven-member solutions, as well. Perhaps, it is not necessary to specify them because the reader can easily construct examples, based on the food-web, shown in Fig.~\ref{foodweb}.

As we noted, there are three parameters in our model and in the following we systematically scan the whole parameter field to identify the dominant solution for each combination of parameters. For this reason, we executed Monte Carlo simulations where a Monte Carlo step ($MCS$) means that on average every node has a chance to update its state. The linear system size varied between $L=100$ to $L=3200$ and the necessary relaxation steps were between $10^3$ to $3\times10^5$ $MCS$s. According to the standard protocol, simulations were launched from an initial state where species are distributed randomly on the lattice and monitor the $\rho_i$ concentration of $X_i$ for all species. We, however, would like to stress that this initial state does not always could be a good choice to find the solution which is valid in the large size limit. It is because a small system size does not necessarily ``offer'' equal chance for all possible solutions to emerge. Some solution, which involves large typical lengths, would emerge just later, but to that stage of the evolutionary process other solutions may invade the whole available space. Therefore, a more complex solution practically has no chance to emerge at a small system size. As a consequence, other solutions may win and an independent run can easily terminates onto a different solution, no matter we use the same set of model parameters. To overcome this uncertainty, we not simply enlarged the system size, but also used alternative initial states where larger and homogeneous domains of competing species were distributed randomly on the lattice. Similar approach was used previously in Refs.\cite{szolnoki_pre11,szolnoki_njp18b}, for instance. In this way, we can create all types of interfaces which could be the foundation to build more sophisticated solutions. To make our analysis more complete, at the critical values of control parameters we also generated sub-solutions independently in restricted areas of space and after we let them fight directly along the separating interface \cite{szolnoki_epl15,szolnoki_njp16,szolnoki_pa18}. This technique can identify the dominant solution unambiguously. The details of the above mentioned protocols will be specified in the next section.

\section{Results}

\subsection{Phase diagrams}

We first summarize our main findings and later we give further details about the microscopic mechanisms which are responsible how dominant solutions emerge. According to our simulations, we can distinguish three main cases which determine the system behavior. The decisive condition is the intensity of interaction between the quartets mentioned in Sec.~\ref{def}. If this interaction is weak, means the value of $\gamma$ is small, then the $A^4_{\{1,3,5,7\}}$ solution dominates the whole $\beta-\alpha$ parameter plane. In other words, only the quartet of ``1+3+5+7'' survive independently of the $\alpha$ and $\beta$ values.
 
\begin{figure}[h!]
	\centering
	\includegraphics[width=7.5cm]{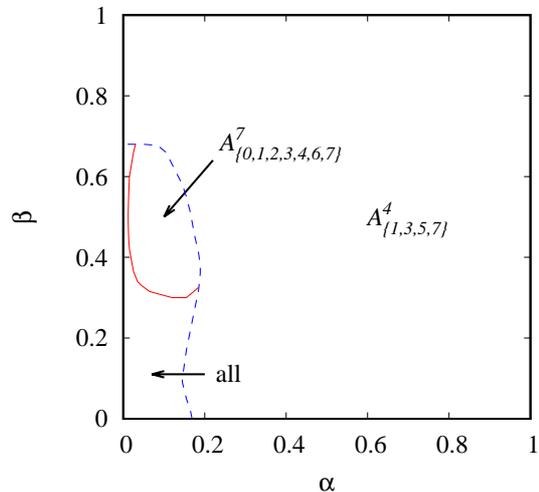}\\
	\caption{Phase diagram on $\beta-\alpha$ parameter plane obtained at $\gamma=0.5$. If $\alpha$ is large enough, the quartet of ``1+3+5+7'' species always win. If both $\beta$ and $\alpha$ are small, all species survive. Interestingly, for low $\alpha$ and intermediate $\beta$ a seven-member solution dominates where species ``5'' is missing. Dashed blue line denotes discontinuous, while solid red line marks the positions of continuous phase transition points.}\label{phd_g05}
\end{figure}

For intermediate values of $\gamma$, when the invasion flow in the large cycle becomes substantial, a conceptually new behavior emerges. We represent this phenomenon by showing the phase diagram obtained at $\gamma=0.5$. Figure~\ref{phd_g05} shows that beside the mentioned $A^4_{\{1,3,5,7\}}$ quartet, other solutions become dominant at certain values of $\beta-\alpha$ pairs. When both $\alpha$ and $\beta$ are small then the ``large circle'' is the winner, hence all eight species coexist. For intermediate $\beta$ values this solution is replaced by $A^7_{\{0,1,2,3,4,6,7\}}$ where only species ``5'' is missing.
 
\begin{figure}[h!]
	\centering
	\includegraphics[width=7.5cm]{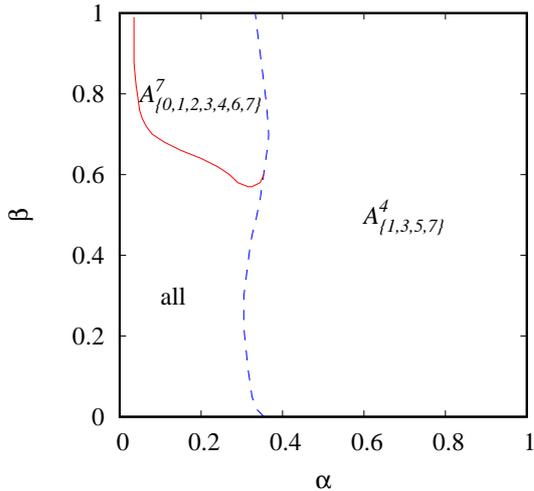}\\
	\caption{Phase diagram on $\beta-\alpha$ parameter plane obtained at $\gamma=1$. The diagram is similar to the one shown in Fig.~\ref{phd_g05}. The only difference is the eight-member full solution occupies larger area in the parameter field, while seven-member solution is shifted toward higher $\beta$ values.}\label{phd_g1}
\end{figure}

Qualitatively similar system behavior can be observed when $\gamma$ is large, hence the flow in the external loop is intensive. In other words, the interaction between the quartets becomes large. This is illustrated in Fig.~\ref{phd_g1} where we present the complete phase diagram on the $\beta-\alpha$ plane. The only difference between the diagrams is the area, where complete eight-species solution dominates, is larger and the seven-member solution is shifted toward larger $\beta$ values.
 
\begin{figure}
	\centering
	\includegraphics[width=7.8cm]{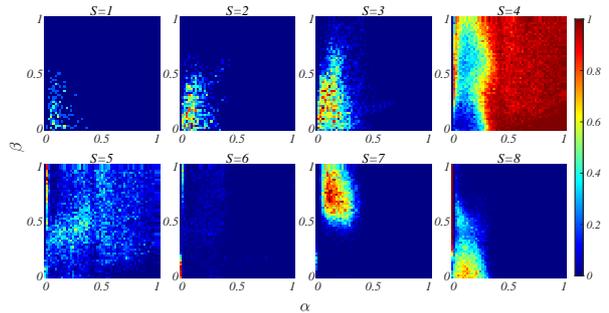}\\
	\caption{Heat maps of the survival probability on $\beta-\alpha$ parameter plane obtained for $\gamma=1$ by using $100 \times 100$ system size. Each panel shows the probability to reach a state which contains $S$ species after $N=L^2$ $MCS$s. The value of $S$ is indicated on each panel. The results are averaged over 100 independent runs.}\label{L1_destinations}
\end{figure}

We must stress that the presented phase diagrams are valid in the large system size limit. If the system size is small then we may observe that the system can easily terminate onto different solutions. To illustrate it, in Fig.~\ref{L1_destinations} we present the survival probabilities at each $(\alpha,\beta)$ parameter pairs for $\gamma=1$, when the linear system size was $L=100$. More precisely, we launched the evolution from a random initial state and monitored the fractions of species until $N= L \times L$ $MCS$s. After, we recorded the number of surviving species and repeated the whole process 100 times. In this way we can estimate the probability that $S$ species survive for long time. The panels of Fig.~\ref{L1_destinations} show these surviving probabilities for all possible $S$ values at
$x$ 
different $(\alpha,\beta)$ pairs on the parameter plane. These panels indicate that different solutions may emerge no matter we use the same values of $(\alpha,\beta)$ parameters. Therefore, the destination is rarely unambiguous, hence the surviving probability equals to 1 only for $S=4$ when $\alpha$ is high and $\beta$ is low. Nevertheless, the contour of areas around ``maximum'' values are roughly agree with the diagram shown in Fig.~\ref{phd_g1}. But, as Fig.~\ref{L1_destinations} illustrated, this system size is insufficient to make reliable conclusions about the proper system behavior.
 
\subsection{Detecting phase transitions}
 
\begin{figure}
	\centering
	\includegraphics[width=7.5cm]{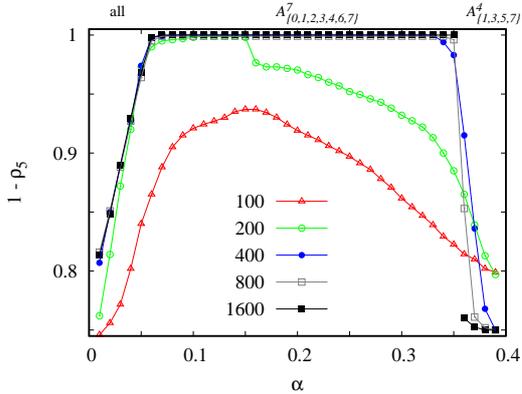}\\
	\caption{The absence of species ``5'', as an order parameter, in dependence of $\alpha$ at $\gamma=1$, $\beta=0.75$ for different system sizes. The linear sizes are indicated in the legend. The dominant solutions are marked on the top. The plots are the average of 100-2000 runs depending on the system size. Lines are just to guide the eye.}\label{A7}
\end{figure}

Therefore, to detect the phase transition points more precisely, we need to apply systematic finite-size analysis. An example is given in Fig.~\ref{A7} where we present the ``lack'' of species ``5'' in dependence of $\alpha$ at fixed $\beta=0.75$ and $\gamma=1$ values. When the value of $1-\rho_5$ reaches 1 then the system enters onto the mentioned $A^7_{\{0,1,2,3,4,6,7\}}$ solution. Evidently, we also checked the portions of other species, because $\rho_5=0$ is fulfilled in other solutions, too. Here the symbols are the average of many independent runs. As an example, for $L=100$ we executed 2000 times. Importantly, the average hides the proper system behavior for small system size, because it mixes different destinations. For example, at $L=100$, $\alpha=0.15$ we never measured $\rho_5=0.063$. Instead, the majority of independent runs terminated onto a state where $\rho_5=0$ or $\rho_5=0.25$. This ambiguity disappears for large system sizes. The plot also highlights that the transition from the eight-member to seven-member solution is continuous because species ``5'' vanishes gradually as we increase $\alpha$. The transition, however, between $A^7_{\{0,1,2,3,4,6,7\}}$ and $A^4_{\{1,3,5,7\}}$ phases is discontinuous. 
 
\begin{figure}
	\centering
	\includegraphics[width=7.5cm]{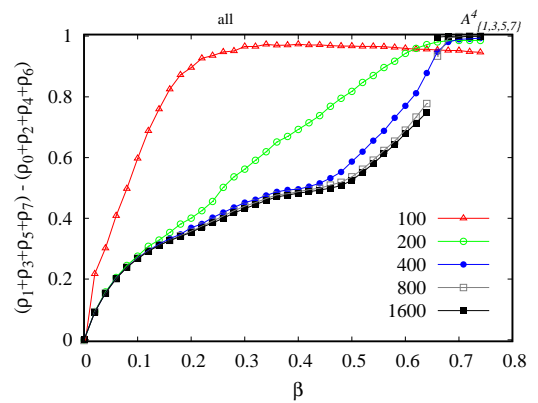}\\
	\caption{The difference between symmetric quartets in dependence of $\beta$ at $\gamma=0.5$, $\alpha=0.1$. When this order parameter becomes 1, the system enters to a four-species state. The dominant solutions are marked on the top. The plots are the average over 20-2000 runs, depending on the system size. The linear size of squares are indicated.}\label{8to4}
\end{figure}

To illustrate another type of phase transition, in Fig.~\ref{8to4} we show an alternative order parameter in dependence of $\beta$ at $\gamma=0.5$ and $\alpha=0.01$. Here, we calculate the difference between the portions of quartets formed by odd- and even-indexed species, respectively. When $\beta$ is small, all available species coexist, the system is in the ``all'' phase, hence the mentioned difference is small. When this parameter becomes 1, then species with even indexes vanish and the quartet of ``1+3+5+7'' species becomes dominant. Interestingly, this formation is viable even if the inner invasion, due to the tiny $\alpha=0.01$, is extremely slow. The explanation of this surprising behavior is given in the next subsection.

Similarly to the previously discussed Fig.~\ref{A7} case, the average of the order parameter could be misleading when the system size is small. This can be clearly seen for $L=100$ and $L=200$, where the average is larger than the value for larger system sizes. It is because the system can easily be trapped in the $A^4_{\{1,3,5,7\}}$ state already at small $\beta$ values. The possible destinations, however, are less ambiguous for larger sizes, but the jump in the order parameter is valid signaling a discontinuous phase transition at $\beta=0.66$.
 
\subsection{Battles of solutions}

In the following, we analyze the possible mechanisms which explain the system behavior summarized in Fig.~\ref{phd_g05} and in Fig.~\ref{phd_g1}. To get a deeper insight about the dominant processes, which drive the patter formation, it is instructive to use a specific initial state where we divide the available space into two halves and both regions are occupied by one of the main quartets formed by odd- or even-indexed species. In this way, we can follow how these solutions compete, or how new possibilities may emerge due to their interaction.

\begin{figure}
	\centering
	\includegraphics[width=3.8cm]{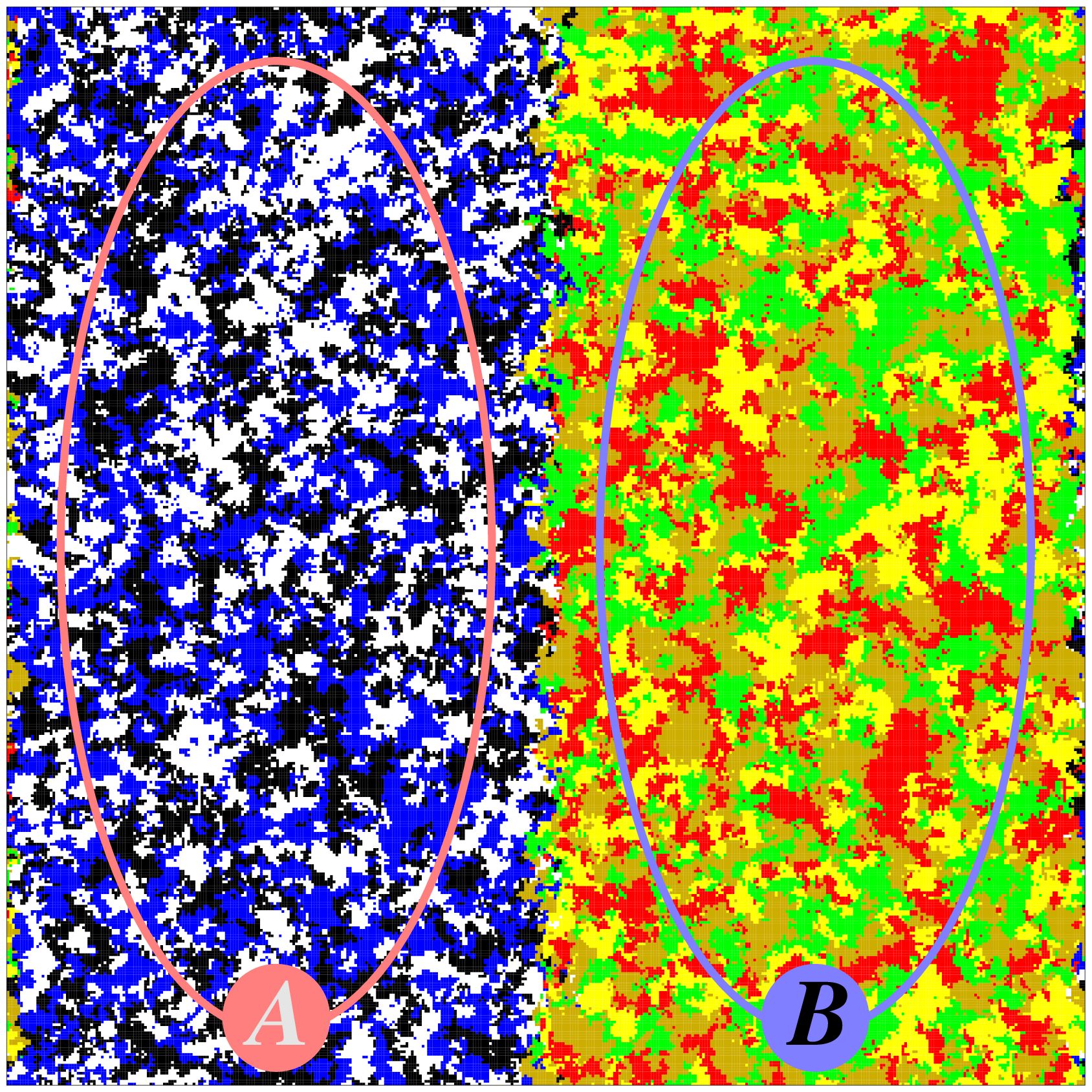}\hspace{0.1cm}\includegraphics[width=3.8cm]{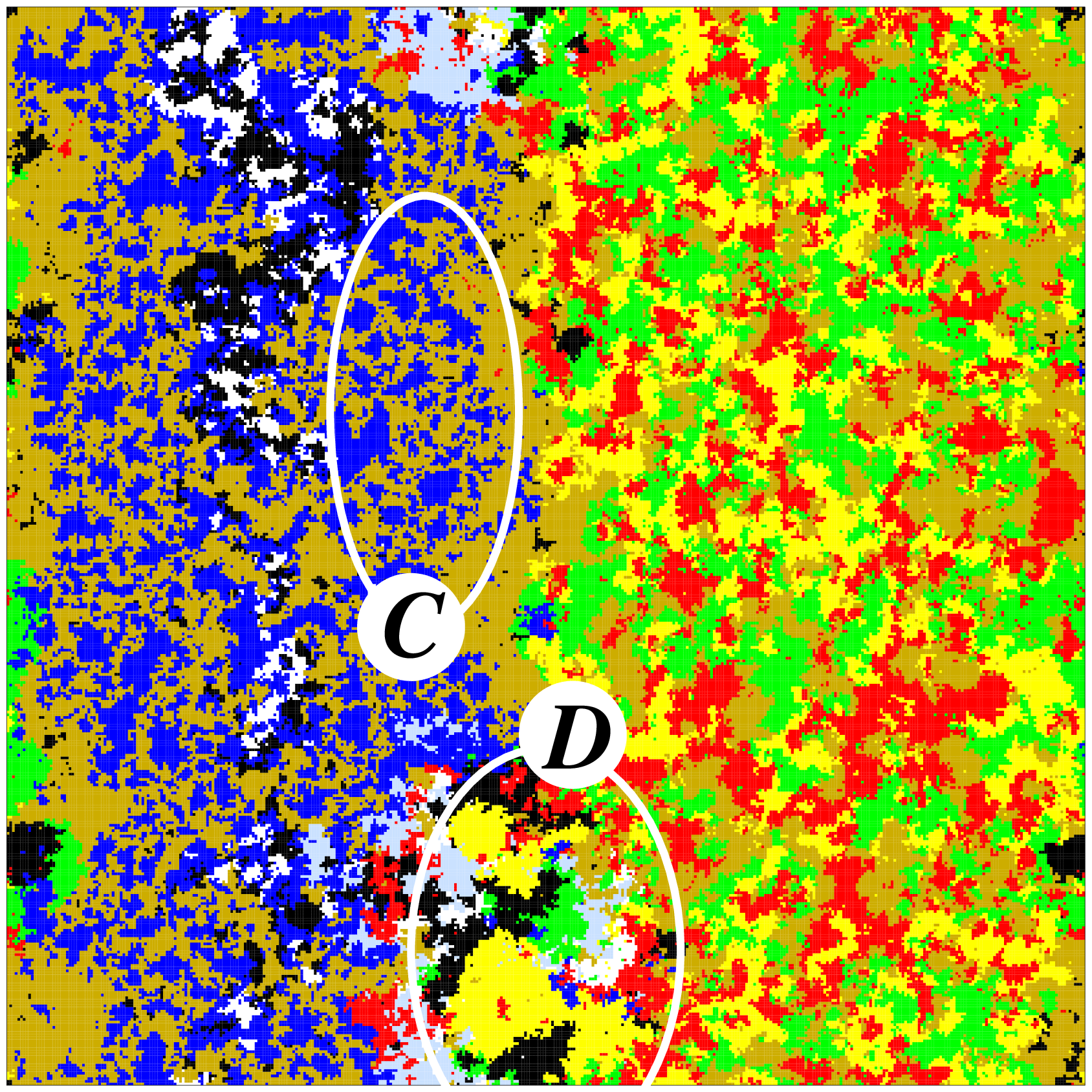}\\
	\caption{Pattern formation in the low $\alpha$ -- low $\beta$ region when we launch the evolution from a prepared state where left (right) side of the space is occupied by even-indexed (odd-indexed) species. In the former case only the $A^3_{\{0,2,4\}}$triplet survive, shown by ``$A$'' on the left panel. For small $\gamma$, when the interaction is weak around the main loop, the $A^4_{\{1,3,5,7\}}$ quartet, marked by ``$B$'', can beat the other solution and gradually invades the whole space. If $\gamma$ is high enough, species ``7'' can easily invade the triplet leaving a neutral $A^2_{\{2,7\}}$ pair behind. But this solution, marked by ``$C$'', cannot survive because the interface between the original quartets is a birthplace of the complete eight-species solution. This is marked by ``$D$''. Since $\gamma$ is high enough, this solution is viable and prevails. Color codes agree with those we introduced in Fig.~\ref{foodweb}. Parameters are: $\alpha=0.1$, $\beta=0.1$, $\gamma=0.05$ (left panel), and $\gamma=1$ (right panel).}\label{small_a_small_b}
\end{figure}

The $\beta-\alpha$ parameter plane can be divided into four main sectors which fundamentally determine the relation of different solutions. We first consider the low $\alpha$ -- low $\beta$ region. It is important to note that the $A^4_{\{0,2,4,6\}}$ quartet formed by even-indexed species is not a proper solution here. Instead, we can see the battle of $A^3_{\{0,2,4\}}$ and $A^3_{\{0,2,6\}}$ triplets when even-indexed species are present. Since species ``4'' beats species ``6'', this fight ends up with the victory of the former solution. This domain is marked by ``$A$'' on the left panel of Fig.~\ref{small_a_small_b}. When $\gamma$ is small then there is just a weak interaction between odd- and even-indexed species. Therefore, the mentioned domains remain compact and they fight along the separating interface. Finally, $A^4_{\{1,3,5,7\}}$ quartet win this battle and the whole space will be occupied by the domain, marked by ``$B$'' on the left panel. 

If $\gamma$ is high enough then the evolution changes drastically. This is illustrated on the right panel of Fig.~\ref{small_a_small_b}. Because of high $\gamma$, species ``7'', who has no predator in the ``0+2+4'' triplet, can easily enter into the $A^3_{\{0,2,4\}}$ domain. The raid of species ``7'' results in a neutral duo. This $A^2{\{2,7\}}$ solution is marked by ``$C$'' on this panel. Interestingly, however, this solution has limited life, because the intensive interactions of the original quartets at the border offer a chance for the complete eight-member solution to emerge. This solution is marked by ``$D$'' on this panel. Because of the high $\gamma$ value, the general invasion flow is intensive in the largest loop, which makes it strong against $A^4_{\{1,3,5,7\}}$. Notably, the  latter quartet is weak due to small $\alpha$. The above described mechanism explains the left-down corners of phase diagrams shown in Fig.~\ref{phd_g05} and in Fig.~\ref{phd_g1}.

We can face a conceptually different situation in the large $\alpha$ -- small $\beta$ corner of the parameter plane, because it provides supporting conditions both for $A^4_{\{0,2,4,6\}}$ and $A^4_{\{1,3,5,7\}}$ quartets. Therefore, both solutions would be viable in the absence of the other. Importantly, the high $\alpha$ value makes the difference from the previously discussed cases, which generates a fast rotation withing both quartets. There is, however, a crucial difference between these solutions: while $A^4_{\{1,3,5,7\}}$ is formed by equal partners, the extra inner invasions described by $\beta$ probability break this delicate balance for the benefit of species ``0'' at the expense of species ``6''. In this way the pattern formed by even-indexed species becomes heterogeneous, including larger domains of ``0'' species. This makes the whole alliance vulnerable against the attack of the rival well-balanced loop.

\begin{figure}
	\centering
	\includegraphics[width=7.5cm]{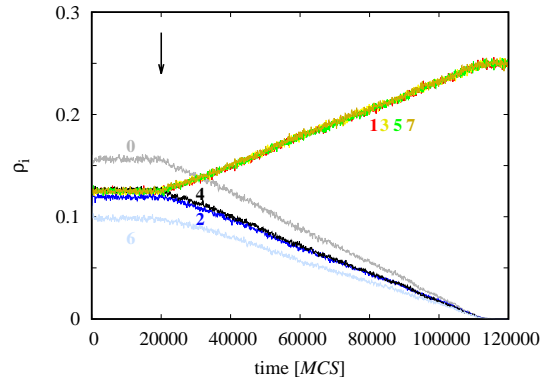}\\
	\caption{The time evolution of species when quartets are fighting at high $\alpha$, low $\beta$ values. Initially both $A^4_{\{0,2,4,6\}}$ and $A^4_{\{1,3,5,7\}}$ solutions evolve independently in separated areas of available space. When separating borders opened, marked by an arrow, symmetric quartet gradually crowd out the rival gang. Color codes are the usual, except white line is replaced by grey one. Parameters are: $\alpha=0.5$, $\beta=0.1$, $\gamma=0.1$, $L=1600$.}\label{4_4}
\end{figure}

This process is illustrated in Fig.~\ref{4_4}, where we monitor the battle of these quartets. Initially, we allowed both solutions to emerge peacefully in restricted areas and the stationary portions of species ``0'' and ``6'' change, as we described previously. Let us stress that in the initial state all species represented equally, which is practically invisible because the unequal stationary distributions of ``0+2+4+6'' species evolve very fast.
After 20000 $MCS$s, when the separating borders opened, the symmetric solution gradually invades the whole space. Notably, the presented simulation was recorded at $\gamma=0.1$, where the interaction between the quartets is moderate. Still, this light communication is capable to reveal the advantage of $A^4_{\{1,3,5,7\}}$ solution. If we apply larger $\gamma$ values then nothing changes conceptually, but the victory of $A^4_{\{1,3,5,7\}}$ becomes faster. Hence, in agreement with the phase diagrams, the symmetric four-member solutions is always the winner in the mentioned corner of the $\beta-\alpha$ parameter plane. 

The lastly discussed observation also answers one of our original questions. Namely, one may argue that the introduction of an additional inner invasion within $A^4_{\{0,2,4,6\}}$ solution results in a faster rotation among these species, which could be useful them \cite{perc_pre07b}. But the weakening consequence of symmetry breaking is stronger, as it was also the case for three-member loops \cite{blahota_epl20}.
 
\begin{figure}[h!]
	\centering
	\includegraphics[width=7.5cm]{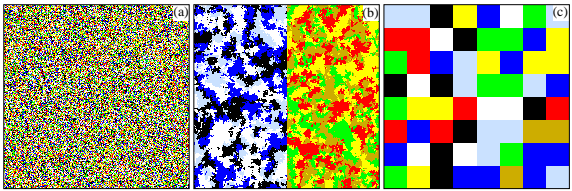}\\
	\caption{Alternative initial states from where evolution is launched when $L=200$. Panel~(a) shows the traditional starting state where every node is uploaded by a randomly chosen species. Panel~(b) shows a state where two competing quartets are generated first. Panel~(c) shows a state where larger patches of species are distributed randomly. The solution, which is valid in the large size limit, can be reached from all starting points, but the necessary system size could be largely different.}\label{initial}
\end{figure}

Next, we discuss the small $\alpha$ -- large $\beta$ corner of the parameter plane. Here, the situation is partly similar to the first discussed case. More precisely, $A^4_{\{0,2,4,6\}}$ solution is unstable and replaced very soon by $A^3_{\{0,2,4\}}$. But this triplet is vulnerable, because the large heterogeneity of inner flow, ($\alpha \ll \beta$), results in huge homogeneous domains, as it was already reported in \cite{juul_pre12} for traditional rock-scissors-paper game. Such a large homogeneous domain is also an easy prey of the rival $A^4_{\{1,3,5,7\}}$ quartet. This picture is valid for small and medium $\gamma$ values, where the interaction between the even- and odd-indexed species is not too strong. For large $\gamma$ values, however, the evolution could be more complex, which can be observed more easily from a ``random-patch'' initial state. This starting pattern is illustrated in Fig.~\ref{initial}. As we already mentioned in Sec.~\ref{def}, this prepared initial state can offer all possible interfaces to be present at the very beginning, which is extremely useful when there are large difference among the invasion rates. In this way, we can observe the valid solution already at smaller system size. We stress, however, that the mentioned state should be reached from all kind of initial states if all available species are present and the system size is large enough. 
 
\begin{figure}
	\centering
	\includegraphics[width=7.5cm]{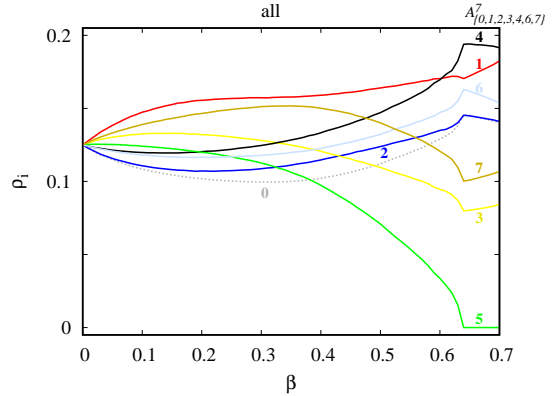}\\
	\caption{Stationary fractions of species in dependence of $\beta$ at $\gamma=1$, $\alpha=0.2$. The results are obtained at $L=3200$ system size. Species are denoted to every lines. For clarity we only show the lines without symbols here. The stable phases are shown on the top.}\label{beta}
\end{figure}

Turning back to the large $\gamma$, large $\beta$, small $\alpha$ region, the above specified prepared initial state can help us to identify the valid $A^7_{\{0,1,2,3,4,6,7\}}$ solution already at relatively small system sizes. At first sight, this solution may seem weird or exotic, but its emergence can be understood if we follow how the portion of species change by increasing $\beta$. This phenomenon is illustrated in Fig.~\ref{beta} where we present the stationary fractions of all species when we increase the intensity of additional invasions at fixed $\gamma=1$ and $\alpha=0.2$. At $\beta=0$ we have a completely symmetric food-web where there are equally strong quartets. Consequently, all species are present at the same level here. When we introduce a non-zero $\beta$ then not only the fractions of species ``0'' and ``6'' start decaying, but simultaneously, the portions of their principal preys, species ``1'' and ``7'' start growing. This increment involves the decay of their preys, which are species ``2'' and ``0''. This double stress explains why species ``0'' suffers the most at small $\beta$ values. Naturally, the decay of species ``6'' affects the population of its principal predator, hence the portion of species ``5'' also a decreasing function. One may argue that the decay of species ``0'' should affect its main predator species ``7''. But the predator of the latter, which is species ``6'', cannot grow here, hence in sum species ``7'' should not necessarily decrease. This difference between the status of species ``5'' and ``7'' explains why there are diverse consequences when both of their principal preys, species ``6'' and ``0'', are attacked directly via an inner invasion described by parameter $\beta$. Importantly, the direct support of species ``4'' via the intensive ``4''$\to$``2'' helps the spreading of species ``4''. This process is also dangerous for species ``5''. On the other hand, the intensive ``2''$\to$``6'' process will not simply strengthen species ``2'', but also weaken its prey species ``3'', which consequence is also enjoyed by species ``4''. When the latter species die out, the remaining seven species form a heterogeneous seven-member Lotka-Volterra loop where the ``weakest" species (species ``4'', who beats species ``6'' with probability $\alpha$) occupies the largest fraction of space. This effect is an example for the phenomenon observed first by Tainaka in the simplest three-member loop \cite{tainaka_pla93}. 

\begin{figure}
	\centering
	\includegraphics[width=7.5cm]{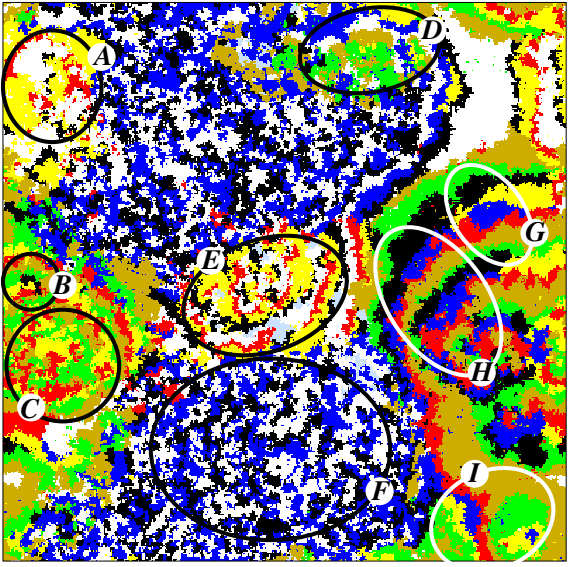}\\
	\caption{Intermediate stage of the evolution taken from the battlefield when different solutions emerge and fight for space. The domains mark the following solutions: $A^4_{\{0,1,3,4\}}$ ($A$), $A^5_{\{1,3,4,5,7\}}$ ($B$), $A^4_{\{1,3,5,7\}}$ ($C$), $A^5_{\{0,2,3,5,7\}}$ ($D$), $A^5_{\{0,1,3,4,6\}}$ ($E$), $A^3_{\{0,2,4\}}$ ($F$), $A^6_{\{1,2,3,4,5,7\}}$ ($G$), $A^5_{\{1,2,4,5,7\}}$ ($H$), and $A^5_{\{1,2,3,5,7\}}$ ($I$). Finally ``$C$'' domain wins. Parameters are $\alpha=0.9$, $\beta=0.9$, $\gamma=1$, and $L=400$.}\label{battlefield}
\end{figure}

Last, we discuss the remaining large $\alpha$ -- large $\beta$ corner of the parameter plane. Here the $A^4_{\{0,2,4,6\}}$ solution is not stable again, hence the remaining $A^3_{\{0,2,4\}}$ solution fights against $A^4_{\{1,3,5,7\}}$ quartet. When $\gamma$ is low and the interaction is weak between these groups then the latter formation wins. This evolution is similar to the case we reported in the left panel of Fig.~\ref{small_a_small_b}. Practically, the same happens when $\gamma$ is high, but in this case a large set of solutions may emerge temporarily. Despite of this diversity, however, the winning alliance remains the symmetric quartet. To give an impression about the variety of different candidates, we present an intermediate snapshot about the ``battlefield'' taken at $\alpha=0.9$, $\beta=0.9$, and $\gamma=1$. Figure~\ref{battlefield} shows that in the intermediate stage of the evolution at least nine(!) solutions emerge locally and fight for space. But eventually the symmetric $A^4_{\{1,3,5,7\}}$ solution crowds out all other competitors.
  
\section{Conclusions}

Which are the most important features of a cyclically dominant alliance that determine its viability against alternative formations? Motivated by this question, we introduced an eight-species model where there are two four-member quartets who interact each other, hence forming a complete eight-member loop, too. The original model is completely symmetric where the inner invasion within the quartets are equally strong. Additionally, we break the symmetry and introduced an extra inner invasion within one of the loops. Based on previous observations about three-member loops, one may argue that this faster rotation of alliance members could be positive to make them stronger. From the other viewpoint, the broken symmetry is always detrimental, hence the new opportunity would just weaken the involved quartet. It is also worth noting that the slight alteration of the original ``symmetric'' food-web offers the chance several potential solutions to emerge. Indeed, an armada of candidates can be observed at intermediate stage of the evolution, but it is believed that the final pattern is determined by some basic concepts.

According to our findings, keeping the symmetry is vital and the solutions, which maintain a balance among their members, are fitter. In the complete parameter space we studied, it can happen that there are more than one solution which possess this attractive character. In this case the general speed of inner invasion could be a decisive factor. If the rotations are equally strong then we could give examples when a short or a larger loop is the winner. For example, a quartet can beat a trio, but a quartet can also beat an octet. Therefore, it cannot be made a simple conclusion that a smaller or larger alliance is better. Probably, there are two competing effects whose relation determines the actual fitness of a loop. On one hand, a short loop may involve relatively large homogeneous domains, which could always be dangerous. On the other hand, too large loop also means that the reaction of the alliance could be delayed, because several inner invasions should happen to produce the predator of the external intruder. Therefore, the sum of these adverse impacts could be case-sensitive.

Naturally, our present study is a sterile theoretical model, but we strongly believe that the observations we made could be useful when real systems are studied. Our another important message is the importance of appropriately chosen system size, which is always a central issue when cyclic dominance is present. Otherwise, the observations could be diverse without deeper understanding. This is why we should always consider the actual size when we want to give predictions about the pattern formation in a finite system driven by intransitive interactions.  

\vspace{0.5cm}

This work was supported by the National Research Foundation of Korea (NRF) grant funded by the Korea government
(MSIT) (No. NRF-2021R1A4A1032924). J.P. was also supported by a grant from Kyung Hee University in 2022 (KHU-20220901). X.C. was supported by the National Natural Science Foundation of China (Grants Nos. 61976048 and 62036002) and the Fundamental Research Funds of the Central Universities of China. A.S. was supported by the National Research, Development and Innovation Office (NKFIH) under Grant No. K142948.

\bibliographystyle{elsarticle-num-names}

\end{document}